\documentclass[11pt,leqno]{article}

\usepackage{amsmath}
\usepackage{amssymb}
\usepackage{amsthm}
\usepackage{array}
\usepackage{authblk}
\usepackage[backend=bibtex]{biblatex}
\usepackage{bm}
\usepackage{caption}
\usepackage{enumitem}
\usepackage{float}
\usepackage{fullpage}
\usepackage[top=1in, bottom=1in, left=1in, right=1in]{geometry}
\usepackage{tikz}
\usepackage{graphicx}
\usepackage[utf8]{inputenc}
\usepackage{mathtools,etoolbox}
\usepackage{setspace}
\usepackage{subcaption}
\usepackage{url}

\title{A Refutation of the Clique-Based P=NP Proofs of LaPlante and Tamta-Pande-Dhami}

\author{Hector A. Cardenas, Chester Holtz, Maria Janczak, Philip Meyers and Nathaniel S. Potrepka}

\affil{Department of Computer Science \\ University of Rochester \\ Rochester, NY 14627, USA}

\date{April 26, 2015}

\begin{document}
\sloppy

{
\singlespacing
\maketitle
}

\begin{abstract}
\noindent In this work, we critique two papers, ``A Polynomial-Time Solution to the Clique Problem'' by Tamta, Pande, and Dhami \cite{tamta}, and ``A Polynomial-Time Algorithm For Solving Clique Problems'' by LaPlante \cite{laplante}. We summarize and analyze both papers, noting that the algorithms presented in both papers are flawed. We conclude that neither author has successfully established that P = NP.
\end{abstract}

\section{Introduction}

Tamta, Pande, and Dhami \cite{tamta} present claimed polynomial-time algorithms for the $k$-clique decision problem and the maximum clique problem, both defined below. They provide a C program in their paper demonstrating their approach.

LaPlante \cite{laplante} presents a single algorithm that attempts to solve the $k$-clique decision problem, maximum clique problem, and the problem of complete enumeration of maximal cliques in polynomial-time. He provides a Java program to demonstrate his algorithm.

In this paper we present background for the clique problems. Then we analyze and critique the algorithms presented in both papers to show that they fail to establish that P = NP.

\subsection{Definitions and Terminology}

First, we will go over some basic terms used throughout the paper. We write P, NP, NP-hard, and NP-complete to describe well-known classes of languages involved in computational complexity theory, a branch of computer science. In particular, P describes the set of languages that can be decided in polynomial-time by a deterministic Turing machine. NP describes the set of languages for which a given solution can be verified in polynomial-time by a deterministic Turing machine. NP-hard describes the set of languages to which any NP problem can be reduced in polynomial-time. NP-complete describes the set of languages that are both NP-hard and in NP.

In graph theory, there are also sets of conventional symbols, function names, and variable definitions that we will briefly cover. Vertex and node are used interchangeably and denote a fundamental unit of the graph. It is important to note that a vertex may exist in a graph yet not be connected to an edge while an edge or arc is, by definition, a link between two vertices. In particular, a graph is defined as a set $V$ of vertices and a set $E$ of edges. We denote the number of vertices and edges as $|V|$ and $|E|$, respectively, and also refer to the number of edges connected to a vertex $v$ as the degree of $v$, denoted $\deg(v)$.

\begin{figure}[H]
	\centering
		\begin{tikzpicture}
          [scale=.8,auto=left, every node/.style={circle,fill=blue!20}]
          \node (n6) at (1,10) {6};
          \node (n4) at (4,8)  {4};
          \node (n5) at (8,9)  [fill=red] {5};
          \node (n1) at (11,8) [fill=red] {1};
          \node (n2) at (9,6)  [fill=red] {2};
          \node (n3) at (5,5)  {3};

          \foreach \from/\to in {n6/n4,n4/n5,n2/n3,n3/n4}
            \draw (\from) -- (\to);
          \foreach \from/\to in {n1/n2,n2/n5,n1/n2,n5/n1}
          	\draw[red](\from) -- (\to);
        \end{tikzpicture}
	\caption{Cliques of G}
	\label{fig:clique}
\end{figure}
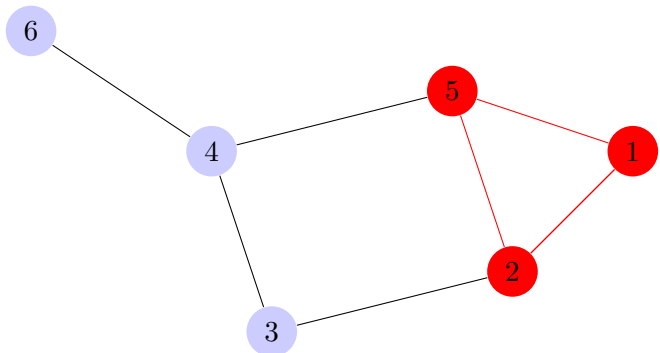

We define a clique as a set of vertices where every pair in the set is connected by an edge. A $3$-clique appears in Figure~\ref{fig:clique}, indicated in red.

The $k$-clique decision problem is concerned with whether a clique containing $k$ vertices exists within a given graph. The maximum clique problem finds a clique with the largest number of vertices. The papers we critique claim to give algorithms solving these problems in polynomial-time.

\section{Tamta, Pande, and Dhami's Argument}

Tamta, Pande, and Dhami \cite{tamta} claim to solve the $k$-clique decision problem in polynomial-time by first reducing it to a maximum flow network interdiction problem (MFNIP) using the method described by Wood \cite{wood}. Tamta, Pande, and Dhami then present a simplification of MFNIP named P-CMFNIP, originally formulated by Wood \cite{wood}, for which they formulate a plausibly polynomial-time algorithm. Tamta, Pande, and Dhami construct an algorithm that supposedly solves the $k$-clique decision problem in polynomial-time. They then derive an algorithm for the maximum clique problem based on their algorithm for the $k$-clique decision problem.

\subsection{Reduction of the $\bm{k}$-clique decision problem to MFNIP}

The maximum flow network interdiction problem (MFNIP,) as described by Wood \cite{wood}, is an NP-complete problem that involves finding the maximum flow through a directed graph from a source node, denoted $s$, to a sink node, denoted $t$ \cite{wood}. Each edge in the graph has an interdiction cost, i.e. the cost of removal. There is also an interdiction budget, which is the maximum allowed total interdiction cost of all the edges removed. The problem is to find the maximum flow through the graph that can be achieved after removing as many edges as possible.

Wood \cite{wood} shows that MFNIP is NP-complete by reducing the $k$-clique decision problem to it. His reduction is as follows. Given an undirected graph $G$, we construct a capacitated directed graph called $G^H$. The construction of $G^H$ from a graph $G = (V, E)$ is given below.
\begin{enumerate}
  \item Let there be a single source node $s$.
  \item Let $N_1$ be a set of vertices containing a vertex $i_e$ for each edge $e \in E$.
  \item Let $N_2$ be a set of vertices containing a vertex $j_v$ for each vertex $v \in V$.
  \item Let there be a single sink node $t$.
  \item For each vertex $i_e \in N_1$, direct an edge with capacity 2 from $s$ to $i_e$, and call the set of these edges $A_1$.
  \item For each edge $e = (u,v)$ in $G$, direct an edge with capacity 1 from $i_e$ to $j_u$, and direct an edge with capacity 1 from $i_e$ to $j_v$. Call the set of these edges $A_2$.
  \item For each vertex $j_v \in N_2$, direct an edge with capacity 1 from $j_v$ to $t$, and call the set of these edges $A_3$.
  \item Let $G^H = (N,A) = (\{s\} \cup \{t\} \cup N_1 \cup N_2, A_1 \cup A_2 \cup A_3)$
\end{enumerate}

Wood \cite{wood} proves that there exists $A_1' \subseteq A_1$ with $|A_1'| = |E| - \binom{k}{2}$ such that the maximum flow from $s$ to $t$ in $G^H - A_1'$ is $k$ if and only if $G$ contains a $k$-clique.

%
%
%

\subsection{Simplification of MFNIP}

Before offering a polynomial-time algorithm, Tamta, Pande, and Dhami first define a simpler problem stemming directly from MFNIP. Tamta, Pande, and Dhami's simplification involves Wood's Lemma 1 \cite{wood}, which states that the maximum flow through $G^H$ constructed from $G$ is equal to the number of vertices $v \in V$ with $\deg(v) > 0$. Consequently, the only way to reduce the maximum flow by removing edges would be to remove all edges connected to a vertex $v$, causing $\deg(v)$ to be $0$. This is equivalent to removing the vertex from the undirected graph altogether. Thus Tamta, Pande, and Dhami's simplification of MFNIP, called P-CMFNIP, is to remove nodes from $N_2$ instead of $N_1$, which is equivalent to removing vertices instead of edges from $G$.
 
\subsection{Tamta, Pande, and Dhami's Claimed Polynomial-Time Algorithm}

Tamta, Pande, and Dhami propose a polynomial-time algorithm to solve the $k$-clique decision problem, which they name ``Poly-Clique.'' Given a graph $G = (V,E)$, the algorithm defines two set-like structures functionally equivalent to priority queues. The queue $T$ stores all vertices $v \in V$, prioritizing the vertices with the lowest interdiction cost, defined as $\deg(v)$. The queue $S$ stores all pairs of vertices $(v_1,v_2)$ that are connected in $G$ by an edge $e \in E$. The pairs are prioritized by the interdiction cost of $v_1$ plus the interdiction cost of $v_2$, minus one, or $\deg(v_1) + \deg(v_2) - 1$. As in Wood's \cite{wood} reduction of the k-clique decision problem to MFNIP, the interdiction budget $R$ is given by the number of edges in $G$ minus $\binom{k}{2}$. Tamta, Pande, and Dhami's algorithm then executes the following procedure.

\newpage

\noindent \textbf{Part 1}
\begin{enumerate}
  \item From $S$, take the vertex pair $(v_1,v_2)$ with the minimum interdiction cost. If $S$ is empty, the behavior of the algorithm is unspecified.
  \item If $\deg(v_1) + \deg(v_2) - 1 \leq R$, continue.  Else, go to step 6.
   \item Let\footnote{Tamta, Pande, and Dhami leave out the addition of $1$ in the third line below. We assume this is a typographical error, since the paper explains that the interdiction cost of a pair of vertices is the sum of their degrees minus $1$.} $$R \leftarrow R - \deg(v_1) - \deg(v_2) + 1$$ $$T \leftarrow T - v_1 - v_2$$ $$S \leftarrow S - (v_1,v_2)$$
  \item Update the priority (interdiction cost) of each vertex in $T$ and each vertex pair in $S$, based on the degrees that result from removing $v_1$ and $v_2$.\footnote{Although Tamta, Pande, and Dhami don't explicitly state this step, we assume it's implied. This step is essential for the algorithm to work properly on virtually any graph.}
  \item Return to step 1.
\end{enumerate}
\textbf{Part 2}
\begin{enumerate}[resume]
  \item From $T$, take the vertex $v$ with the minimum interdiction cost. If $T$ is empty, the behavior of the algorithm is unspecified.
  \item If $\deg(v) \leq R$, continue.  Else, go to step 11.
  \item Let $$R \leftarrow R - \deg(v).$$ $$T \leftarrow T - v$$ and, for all $i$, such that there exists an edge in $E$ connecting $(v,v_i)$ $$S \leftarrow S - (v,v_i)$$
  \item Update the priority (interdiction cost) of each vertex in $T$ and each vertex pair in $S$, based on the degrees that result from removing $v$.
  \item Go back to step 6.
\end{enumerate}
\textbf{Part 3}
\begin{enumerate}[resume]
  \item If $|T| = k$, then we conclude that $G$ contains a clique of size $k$. Otherwise, we conclude that $G$ does not contain a clique of size $k$.
\end{enumerate}

Further, as demonstrated by Wood, a polynomial-time solution to the $k$-clique decision problem implies a polynomial-time solution to the maximum clique problem.

\section{Critique of Tamta, Pande, and Dhami}

In our analysis of Tamta, Pande, and Dhami's \cite{tamta} polynomial-time $k$-clique algorithm, we identified several flaws and inconsistencies as well as a well-defined set of graphs for which Tamta, Pande, and Dhami's algorithm is not guaranteed to succeed in finding the correct maximum clique.

\subsection{A Simple Counterexample to Tamta, Pande, and Dhami's Algorithm}

\begin{figure}[H]
	\centering
		\begin{tikzpicture}
          [scale=.8,auto=left,every node/.style={circle,fill=blue!20}]
          \node (n1) at (1,10) {1};
          \node (n2) at (5,10) {2};
          \node (n3) at (3,9)  {3};
          \node (n4) at (2,7)  [fill=red] {4};
          \node (n5) at (4,7)  [fill=red] {5};
          \node (n6) at (1,5)  [fill=red] {6};
          \node (n7) at (5,5)  [fill=red] {7};

          \foreach \from/\to in {n1/n2,n1/n3,n2/n3,n1/n2,n1/n4,n2/n5,n3/n4,n3/n5,n1/n6,n2/n7}
            \draw (\from) -- (\to);
            
          \foreach \from/\to in {n4/n5,n4/n6,n4/n7,n5/n7,n5/n6,n7/n6}
          	\draw[red](\from) -- (\to);
        \end{tikzpicture}
	\caption{A graph for which Tamta, Pande, and Dhami's algorithm fails when $k = 4$ \cite{wiki:clique_problem}}
	\label{fig:tceI}
\end{figure}
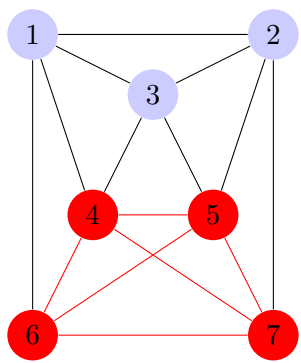

In Figure~\ref{fig:tceI} we present the following $7$-node graph as a simple counterexample to Tamta, Pande, and Dhami's algorithm.\footnote{This graph came from the Wikipedia article \textit{Clique problem} \cite{wiki:clique_problem}} Consider the execution of their algorithm for the $4$-clique decision problem on this graph. This graph contains a $4$-clique, highlighted in red, so we should expect the algorithm to identify that a $4$-clique is present.

As in Wood's \cite{wood} reduction of the $k$-clique decision problem to MFNIP, the graph's interdiction budget, $R$, is given by $R = |E| - \binom{k}{2}$. In this case, $R = 15 - 6 = 9$. The algorithm may therefore remove up to $9$ edges from the graph. As stated above, Tamta, Pande, and Dhami's algorithm begins by finding a connected pair of vertices with a minimal interdiction cost. In the above graph, any pair containing two vertices in the set $\{1,2,6,7\}$ have an interdiction cost of $7$, and no pair of vertices has a lower interdiction cost. Therefore, the algorithm will remove one of those pairs.

Tamta, Pande, and Dhami don't specify the behavior of the algorithm if more than one pair of vertices has a minimal interdiction cost. Suppose the algorithm chooses to interdict the pair $(6,7)$. The figure below shows the result of doing so.

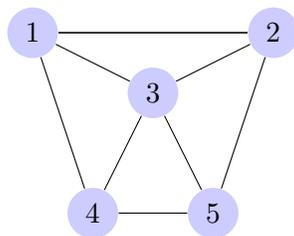
\begin{figure}[H]
	\centering
		\begin{tikzpicture}
          [scale=.8,auto=left,every node/.style={circle,fill=blue!20}]
          \node (n1) at (1,10) {1};
          \node (n2) at (5,10) {2};
          \node (n3) at (3,9)  {3};
          \node (n4) at (2,7)  {4};
          \node (n5) at (4,7)  {5};

          \foreach \from/\to in {n1/n2,n1/n3,n2/n3,n1/n2,n1/n4,n2/n5,n3/n4,n3/n5,n4/n5}
            \draw (\from) -- (\to);
        \end{tikzpicture}
	\caption{Example graph for which Tamta, Pande, and Dhami's algorithm potentially fails for $k = 4$ (with vertices $6$ and $7$ removed}
	\label{fig:tceIReduced}
\end{figure}

The original interdiction budget was $9$, and removing all edges connected to the vertex pair $(6,7)$ had a cost of $7$. Thus the remaining interdiction budget is $2$. But we can't disconnect any vertices from the above graph by removing only $2$ vertices. The algorithm proceeds to the final step, comparing $k$ to the number of elements in the set $T$, i.e. the number of vertices still connected in the graph. $T$ contains $5$ elements, but $k$ equals $4$. The output of the algorithm is that the original graph didn't contain a $4$-clique, contrary to fact.

This example illustrates one of many instances for which Tamta, Pande, and Dhami's algorithm fails. Tamta, Pande, and Dhami's algorithm only succeeds if it doesn't delete any vertices from the largest clique in the graph. They assume that the algorithm won't delete any vertices from the largest clique because these vertices will never have the lowest degree of any vertex in the graph. As we see in this example, as well as in the more general counterexample below, this assumption is flawed.

\subsection{A More General Counterexample to Tamta, Pande, and Dhami's Algorithm}

\begin{figure}[H]
	\centering
		\begin{tikzpicture}
          [scale=.8,auto=left,every node/.style={circle,fill=blue!20}]
          \node (n1) at (3,8) {\ \ $C_k$\ \ \ };
          \node (n2) at (5,10){$C'_{k-1}$};
          \node (n3) at (1,10){$C_{k-1}$};

          \foreach \from/\to in {n1/n2, n1/n3}
            \draw (\from) -- (\to);
          \foreach \from/\to in {n2/n3}
            \draw[black,ultra thick] (\from) -- (\to);
        \end{tikzpicture}
	\caption{General counterexample}
	\label{fig:tceII}
\end{figure}
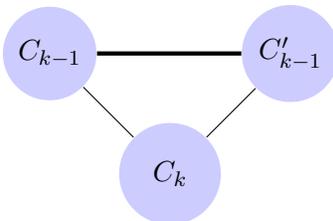

We now demonstrate that Tamta, Pande, and Dhami's algorithm fails in an unlimited set of cases. For every k $\geq 4$, we construct a graph $G_k$ for which Tamta, Pande, and Dhami's algorithm fails. Let $C_k$, $C_{k-1}$, and $C'_{k-1}$ be cliques of size $k$ and $k-1$, respectively. Each vertex in $C_k$ is connected to exactly $k-1$ other vertices in the clique. Likewise, each vertex in $C_{k-1}$ is connected to exactly $k-2$ other vertices in the clique. Consider the graph in Figure~\ref{fig:tceII}.

Let a thin black line denote a single edge from one vertex in $C_k$ to one vertex in $C_{k-1}$ or $C'_{k-1}$. Thus all but two vertices in $C_k$ are connected to exactly $k-1$ other vertices, and two vertices in $C_k$ (those connected to either $C_{k-1}$ or $C'_{k-1}$) are connected to exactly $k$ vertices. Let these two vertices be called $v$ and $v'$.

Let the thick black line denote $k-2$ edges from those $k-2$ vertices in $C_{k-1}$ not connected to $C_k$ to corresponding vertices in $C'_{k-1}$. Thus each vertex of $C_{k-1}$ and $C'_{k-1}$ is connected to exactly $k-1$ vertices.

This construction yields a graph where each pair of adjacent vertices, except those including $v$ or $v'$, have an interdiction cost of exactly $(k-1)+(k-1)-1 = 2k-3$. Those pairs including $v$ or $v'$ have an interdiction cost greater than $2k-3$, since $v$ and $v'$ are connected to more than $k-1$ vertices. Therefore, any pair of adjacent vertices, except those including $v$ or $v'$, have the minimum interdiction cost $2k-3$. Since $k \geq 4$, there exists at least one\footnote{There are $\binom{k - 2}{2}$ such pairs, to be precise} pair of adjacent vertices in $C_k$ that doesn't contain $v$ or $v'$. If this pair is removed, the $k$-clique is destroyed, and the algorithm fails.

\section{LaPlante's Approach}

LaPlante \cite{laplante} proposes a polynomial-time algorithm for solving the $k$-clique decision problem, the $k$-clique enumeration problem, and the maximum clique problem. His approach to each of these clique problems consists of two phases: for each vertex, find every $3$-clique containing it and list all vertices in each of these cliques, then iterate through the list of $3$-cliques to identify larger cliques.

Here we use LaPlante's terminology to describe his approach. For each vertex $v$ in the graph, phase 1 finds all $3$-cliques containing it. LaPlante's algorithm finds these $3$-cliques through a process he calls ``neighbor introductions.'' The idea is based on a network structure where each vertex in the graph works as a node in the network that can ``communicate'' with the vertices it's connected to, its ``neighbors.'' Each of the vertices tell their neighbors of their other neighbors. When a vertex hears of a neighbor of one of its neighbors that it is also a neighbor to, it knows that it is in a $3$-clique with those two vertices. LaPlante claims that finding all $3$-cliques is achieved in O$(n^3)$ time.

After phase 1 is complete, each vertex is associated with a list of all $3$-cliques it is a member of. For any vertex $v$, the set of $3$-cliques that include $v$ is the ``neighborhood'' of $v$. LaPlante devised a method for visualizing the neighborhood of a vertex $v$ by placing the vertex $v$ at the center and depicting each $3$-clique in its neighborhood as a triangle stemming from the central vertex. LaPlante uses the term ``vertex pair'' to refer to a pair of vertices that are part of a $3$-clique together with the central vertex in a neighborhood. We will use this term accordingly in our analysis of LaPlante's algorithm.

phase 2 iterates through each vertex $v$, and looks at the neighborhood of $v$. It then arbitrarily chooses a vertex pair in this neighborhood. The key node is one of the two nodes in the vertex pair also arbitrarily chosen. The algorithm iterates through the other vertex pairs in the neighborhood, looking for a pair containing the key node. If the key node is found, the algorithm checks for all other vertex pairs for which a merge with this vertex pair can occur. For example, if vertex pair $1$, $2$ is chosen as the initial pair and 1 is the key node. If the algorithm finds the pair $1$, $3$, it will need to check for pair $2$, $3$ before merging. After all possible merges are completed, the algorithm loop back, beginning with a vertex pair it has not yet merged until no such vertex pair exists. This process is described in further detail in section 5.

\section{Critique of LaPlante}
\subsection{Counterexample Introduction}
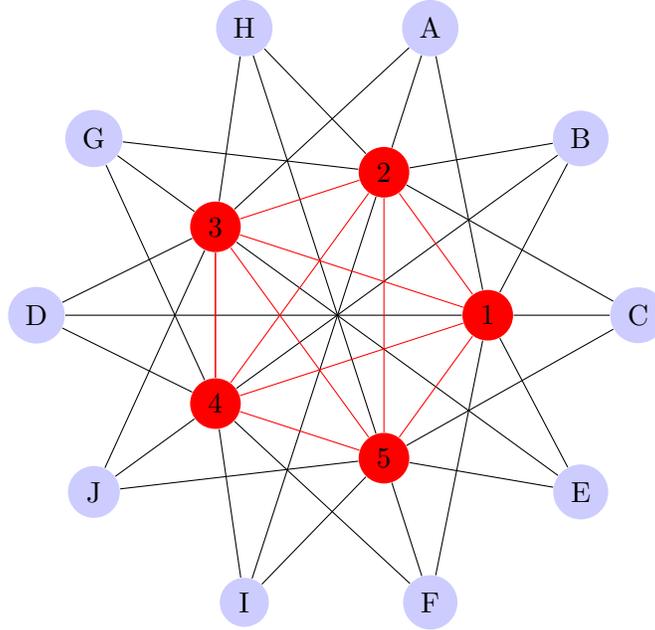
\begin{figure}[H]
	\centering
        \begin{tikzpicture}
		 [scale=.8,auto=left,every node/.style={circle,fill=blue!20}]
          \node (n1) at (2.5,0)         [fill=red] {1};
          \node (n2) at (.775,2.3775)   [fill=red] {2};
          \node (n3) at (-2.025,1.47)   [fill=red] {3};
          \node (n4) at (-2.025,-1.47)  [fill=red] {4};
          \node (n5) at (0.775,-2.3775) [fill=red] {5};
          
          \node (n8) at (5,0)           {C};
          \node (n7) at (4.045,2.939)   {B};
          \node (n6) at (1.545,4.775)   {A};
          \node (n13)at (-1.545,4.775)  {H};
          \node (n12)at (-4.045,2.939)  {G};
          \node (n9) at (-5,0)          {D};
          \node (n15)at (-4.045,-2.939) {J};
          \node (n14)at (-1.545,-4.775) {I};
          \node (n11)at (1.545,-4.775)  {F};
          \node (n10)at (4.045,-2.939)  {E};

		  \foreach \from/\to in {n1/n6,n2/n6/,n3/n6,n1/n7,n2/n7,n4/n7,n1/n8,n2/n8,n5/n8,n1/n9,n3/n9,n4/n9,n1/n10,n3/n10,n5/n10,n1/n11,n4/n11,n5/n11,n2/n12,n3/n12,n4/n12,n2/n13,n3/n13,n5/n13,n2/n14,n4/n14,n5/n14,n3/n15,n4/n15,n5/n15}
            \draw (\from) -- (\to);
          \foreach \from/\to in {n1/n2,n1/n3,n1/n4,n1/n5,n2/n3,n2/n4,n2/n5,n3/n4,n3/n4,n3/n5,n4/n5}
          	\draw[red](\from) -- (\to);
       \end{tikzpicture}
	\caption{Example graph for which LaPlante's algorithm fails}
	\label{fig:lceI}
\end{figure}

It is sufficient to prove that LaPlante's algorithm is invalid if at least one graph is found for which the algorithm fails to find the maximum clique.
Consider the graph displayed in Figure \ref{fig:lceI}.

The graph in Figure~\ref{fig:lceI} contains a $5$-clique containing the vertices $1$, $2$, $3$, $4$, and $5$. It also contains $4$-cliques each consisting of $3$ vertices from the $5$-clique, and another vertex labeled with a letter that isn't part of the $5$-clique. There is one $4$-clique for every combination of three vertices from the $5$ in the central $5$-clique. Note that the graph has $5$-way rotational symmetry. The graph contains vertices on the outside labeled with letters that are only part of a $4$-clique, and those on the inside labeled with numbers that are part of the $5$-clique as well as $6$ $4$-cliques. LaPlante's algorithm works by looking at each vertex and finding the largest clique in which it appears, then noting the largest of the maximal cliques around each vertex. Since each vertex in the above graph belongs to one of two cases, in the analysis below it suffices to demonstrate the result of applying LaPlante's Algorithm on a vertex from each case, without loss of generality. Below we apply LaPlante's algorithm to vertices $1$ and A. The actual algorithm would execute this procedure for each vertex; however, the result for vertices $2$, $3$, $4$, $5$ would be equivalent to the result for $1$, and the results for vertices B, C, D, E, F, G, H, I, J would be equivalent to that of A.

\subsection{Phase 1}
The algorithm will begin by executing phase 1. The execution of phase 1 gives each vertex a list of all $3$-cliques it is a part of. After the completion of this step, each vertex will have ``knowledge'' of its neighborhood, which can be represented by a ring-shaped graph around the vertex in question, as was done by LaPlante in his paper.

Phase 2 follows phase 1, and will execute around every vertex. As was described above, due to symmetry, all vertices are either an outside letter vertex or an inside number vertex, and thus below it is sufficient to describe the execution of phase 2 around one of each.

\subsection{Phase 2 around a Letter Vertex}

Recall a letter vertex in this example is one of the outside vertices which is only part of a maximum size $4$-clique. Here we consider phase 2 around vertex A.

\begin{figure}[H]
	\centering
		\begin{tikzpicture}
          [scale=.8,auto=left,every node/.style={circle,fill=blue!20}]
          \node (n1) at (0,0)   {A};
          \node (n2) at (-.5,3) {1};
          \node (n3) at (.5,3)  {2};
          \node (n4) at (3,.5)  {2};
          \node (n5) at (3,-.5) {3};
          \node (n6) at (.5,-3) {1};
          \node (n7) at (-.5,-3){3};

          \foreach \from/\to in {n1/n2,n1/n3,n1/n4,n1/n5,n1/n6,n1/n7,n2/n3,n4/n5,n6/n7}
            \draw (\from) -- (\to);
        \end{tikzpicture}
	\caption{LaPlante counterexample II}
	\label{fig:lceII}
\end{figure}
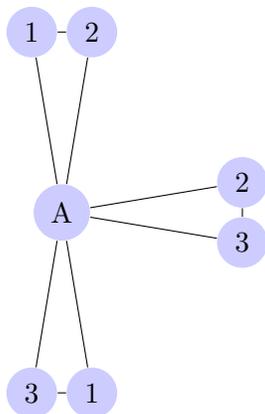

Figure \ref{fig:lceII} depicts the neighborhood of $3$-cliques involving vertex $A$.

The algorithm proceeds to arbitrarily choose a pair of vertices to start with. Recall that a pair is a set of two vertices which along with the central vertex in question compose a $3$-clique. Suppose the algorithm selects the pair $1$ and $2$ and uses node $1$ as the key vertex, which is chosen arbitrarily. The algorithm will proceed to check other pairs looking for vertex $1$, the key, and will locate the pair $\{1, 3\}$. It will then check if vertex pair  $\{2, 3\}$ exists, find it, and then merge vertices $1$, $2$, and $3$ together to create a $4$-clique. After this point, the algorithm will recognize that no more vertices remain, and note that the $4$-clique $A$, $1$, $2$, $3$ is a maximum clique involving $A$. This is correct.

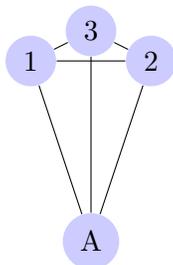
\begin{figure}[H]
	\centering
		\begin{tikzpicture}
          [scale=.8,auto=left,every node/.style={circle,fill=blue!20}]
          \node (n1) at (0,0)  {A};
          \node (n2) at (-1,3) {1};
          \node (n3) at (1,3)  {2};
          \node (n4) at (0,3.5){3};

          \foreach \from/\to in {n1/n2,n1/n3,n1/n4,n2/n3,n2/n4,n3/n4}
            \draw (\from) -- (\to);
        \end{tikzpicture}
	\caption{LaPlante counterexample III}
	\label{fig:lceIII}
\end{figure}

\subsection{Phase 2 around a Number Vertex}

Recall a number vertex in this example is one of the inside vertices which is part of the central $5$-clique and six $4$-cliques. Here we consider phase 2 around vertex $1$.

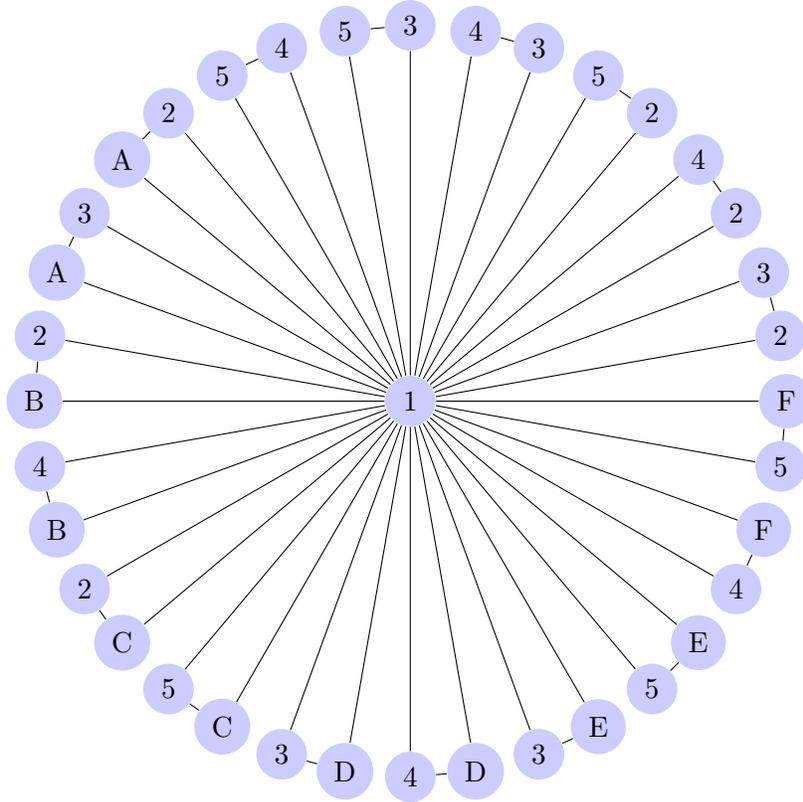
\begin{figure}[H]
	\centering
		\begin{tikzpicture}
          [scale=.5,auto=left,every node/.style={circle,fill=blue!20}]
          \node (n1) at (0,0)                           {1};
          \node (n21) at (9.84807753012,1.73648177667)  {2};
          \node (n31) at (9.39692620786,3.42020143326)  {3};
          \node (n22) at (8.66025403784,5)              {2};
          \node (n41) at (7.66044443119,6.42787609687)  {4};
          \node (n23) at (6.42787609687,7.66044443119)  {2};
          \node (n51) at (5,8.66025403784)              {5};
          \node (n32) at (3.42020143326,9.39692620786)  {3};
          \node (n42) at (1.73648177667,9.84807753012)  {4};
          \node (n33) at (0,10)                         {3};
          \node (n52) at (-1.73648177667,9.84807753012) {5};
          \node (n43) at (-3.42020143326,9.39692620786) {4};
          \node (n53) at (-5,8.66025403784)             {5};
          \node (n24) at (-6.42787609687,7.66044443119) {2};
          \node (na1) at (-7.66044443119,6.42787609687) {A};
          \node (n34) at (-8.66025403784,5)             {3};
          \node (na2) at (-9.39692620786,3.42020143326) {A};
          \node (n25) at (-9.84807753012,1.73648177667) {2};
          \node (nb1) at (-10,0)                        {B};
          \node (n44) at (-9.84807753012,-1.73648177667){4};
          \node (nb2) at (-9.39692620786,-3.42020143326){B};
          \node (n26) at (-8.66025403784,-5)            {2};
          \node (nc1) at (-7.66044443119,-6.42787609687){C};
          \node (n54) at (-6.42787609687,-7.66044443119){5};
          \node (nc2) at (-5,-8.66025403784)            {C};
          \node (n35) at (-3.42020143326,-9.39692620786){3};
          \node (nd1) at (-1.73648177667,-9.84807753012){D};
          \node (n45) at (0,-10)                        {4};
          \node (nd2) at (1.73648177667,-9.84807753012) {D};
          \node (n36) at (3.42020143326,-9.39692620786) {3};
          \node (ne1) at (5,-8.66025403784)             {E};
          \node (n55) at (6.42787609687,-7.66044443119) {5};
          \node (ne2) at (7.66044443119,-6.42787609687) {E};
          \node (n46) at (8.66025403784,-5)             {4};
          \node (nf1) at (9.39692620786,-3.42020143326) {F};
          \node (n56) at (9.84807753012,-1.73648177667) {5};
          \node (nf2) at (10,0)                         {F};

          \foreach \from/\to in {n1/n21,n1/n31,n21/n31,n1/n22,n1/n41,n22/n41,n1/n23,n1/n51,n23/n51,n1/n32,n1/n42,n32/n42,n1/n33,n1/n52,n33/n52,n1/n43,n1/n53,n43/n53,n1/n24,n1/na1,n24/na1,n1/n34,n1/na2,n34/na2,n1/n25,n1/nb1,n25/nb1,n1/n44,n1/nb2,n44/nb2,n1/n26,n1/nc1,n26/nc1,n1/n54,n1/nc2,n54/nc2,n1/n35,n1/nd1,n35/nd1,n1/n45,n1/nd2,n45/nd2,n1/n36,n1/ne1,n36/ne1,n1/n55,n1/ne2,n55/ne2,n1/n46,n1/nf1,n46/nf1,n1/n56,n1/nf2,n56/nf2}
            \draw (\from) -- (\to);
        \end{tikzpicture}
	\caption{LaPlante counterexample IV}
	\label{fig:lceIV}
\end{figure}

Figure \ref{fig:lceIV} depicts the neighborhood of $3$-cliques involving vertex $1$.

LaPlante's algorithm will begin phase 2 by arbitrarily picking one of the vertex pairs in the neighborhood depicted in Figure \ref{fig:lceIV}. In its current state, there only exist two possible types of vertex pairs: pairs that contain vertices denoted by a combination of a letter and a number (an outer and an inner vertex paired together), and pairs where both vertices are designated by numbers (two inner vertices paired together). The algorithm will merge with the chosen vertex pair, and then restart the merging with another vertex pair until all vertex pairs have been merged into a clique at least once. In the subsections to follow, we demonstrate that both choices for a start pair (number-letter or number-number pair) can result in the algorithm failing to discover the $5$-clique. Thus, after iterating through all the possible pairs to start the merge on, the algorithm could still fail to discover the $5$-clique. Therefore, the algorithm fails for this graph.

\subsubsection{Merge Starting with a Number-Letter Pair}

Suppose the algorithm selects a number-letter pair as the vertex pair with which to begin the merge. Since each lettered vertex is a member of a maximal $4$-clique, when the algorithm merges vertex pairs containing a single lettered vertex, it will discover a $4$-clique, and stop merging because all vertex pairs containing that letter would have already been merged into the maximal $4$-clique. Take the example in Figure~\ref{fig:lceV} started with the vertex pair $\{2, A\}$.

\begin{figure}[H]
	\centering
		\begin{tikzpicture}
          [scale=.5,auto=left,every node/.style={circle,fill=blue!20}]
          \node (n1)  at (0,0)                          {1};
          \node (n22) at (8.66025403784,5)              {2};
          \node (n41) at (7.66044443119,6.42787609687)  {4};
          \node (n23) at (6.42787609687,7.66044443119)  {2};
          \node (n51) at (5,8.66025403784)              {5};
          \node (n32) at (3.42020143326,9.39692620786)  {3};
          \node (n42) at (1.73648177667,9.84807753012)  {4};
          \node (n33) at (0,10)                         {3};
          \node (n52) at (-1.73648177667,9.84807753012) {5};
          \node (n43) at (-3.42020143326,9.39692620786) {4};
          \node (n53) at (-5,8.66025403784)             {5};
          \node (n24) at (-6.42787609687,7.66044443119) {2};
          \node (n34) at (-8.19152044289,5.73576436351) {3};
          \node (na2) at (-9.39692620786,3.42020143326) {A};
          \node (n25) at (-9.84807753012,1.73648177667) {2};
          \node (nb1) at (-10,0)                        {B};
          \node (n44) at (-9.84807753012,-1.73648177667){4};
          \node (nb2) at (-9.39692620786,-3.42020143326){B};
          \node (n26) at (-8.66025403784,-5)            {2};
          \node (nc1) at (-7.66044443119,-6.42787609687){C};
          \node (n54) at (-6.42787609687,-7.66044443119){5};
          \node (nc2) at (-5,-8.66025403784)            {C};
          \node (n35) at (-3.42020143326,-9.39692620786){3};
          \node (nd1) at (-1.73648177667,-9.84807753012){D};
          \node (n45) at (0,-10)                        {4};
          \node (nd2) at (1.73648177667,-9.84807753012) {D};
          \node (n36) at (3.42020143326,-9.39692620786) {3};
          \node (ne1) at (5,-8.66025403784)             {E};
          \node (n55) at (6.42787609687,-7.66044443119) {5};
          \node (ne2) at (7.66044443119,-6.42787609687) {E};
          \node (n46) at (8.66025403784,-5)             {4};
          \node (nf1) at (9.39692620786,-3.42020143326) {F};
          \node (n56) at (9.84807753012,-1.73648177667) {5};
          \node (nf2) at (10,0)                         {F};
          
          \foreach \from/\to in {n1/n22,n1/n41,n22/n41,n1/n23,n1/n51,n23/n51,n1/n32,n1/n42,n32/n42,n1/n33,n1/n52,n33/n52,n1/n43,n1/n53,n43/n53,n1/n25,n1/nb1,n25/nb1,n1/n44,n1/nb2,n44/nb2,n1/n26,n1/nc1,n26/nc1,n1/n54,n1/nc2,n54/nc2,n1/n35,n1/nd1,n35/nd1,n1/n45,n1/nd2,n45/nd2,n1/n36,n1/ne1,n36/ne1,n1/n55,n1/ne2,n55/ne2,n1/n46,n1/nf1,n46/nf1,n1/n56,n1/nf2,n56/nf2, n1/n24, n1/n34, n1/na2, n24/n34, n24/na2, na2/n34}
            \draw (\from) -- (\to);
        \end{tikzpicture}
	\caption{LaPlante counterexample V}
	\label{fig:lceV}
\end{figure}

The algorithm recognizes that both vertex pairs $\{2, 3\}$ and $\{3, A\}$ are elements of the set of $3$-cliques surrounding node $1$, so it merges them. The algorithm stops here since there are no more vertex pairs that can be merged into the set, due no more pairs containing the key node A. A $5$ clique is not found.

\subsubsection{Merge Starting with a Number-Number Pair}
In another iteration, the algorithm might choose to begin the merge with a number-number pair. Note that from here the algorithm has two ways to decide a merge. It can either merge with another number-number pair and from there work to find the $5$-clique, or merge with a number-letter pair and from there go to discover a maximum clique of size 4. LaPlante does not specify which merge should happen, but he seems to take it that any merge that creates a larger clique is a merge that can be taken. Thus it cannot be assumed that the algorithm will merge with a number-number pair. Furthermore, since the algorithm never backtracks to this point once the merge has been made, if it first merges with a number-letter pair each time it considers the number-number vertex-pairs surrounding a vertex within the $5$-clique. It will not find the $5$-clique, just all of the $4$-cliques.

For example, if the pair $\{2, 3\}$ was chosen to begin with, it could be merged next with either $\{2, 4\}$, $\{2, 5\}$, $\{3, 4\}$, $\{3,5\}$, \{2, A\}, or \{3, A\}. The choice the algorithm makes is arbitrary, and LaPlante seems to take it that the chosen merge is irrelevant. If $\{2, 4\}$, $\{2, 5\}$, $\{3, 4\}$, or $\{3, 5\}$ are chosen, the algorithm will continue merging to arrive at a $5$-clique.

\begin{figure}[H]
	\centering
		\begin{tikzpicture}
          [scale=.5,auto=left,every node/.style={circle,fill=blue!20}]
          \node (n1)  at (0,0)                          {1};
          \node (n2)  at (9.39692620786,3.42020143326)  {3};
          \node (n3)  at (6.42787609687,7.66044443119)  {2};
          \node (n4)  at (1.73648177667,9.84807753012)  {4};
          \node (n5)  at (-3.42020143326,9.39692620786) {5};
          \node (n24) at (-6.42787609687,7.66044443119) {2};
          \node (na1) at (-7.66044443119,6.42787609687) {A};
          \node (n34) at (-8.66025403784,5)             {3};
          \node (na2) at (-9.39692620786,3.42020143326) {A};
          \node (n25) at (-9.84807753012,1.73648177667) {2};
          \node (nb1) at (-10,0)                        {B};
          \node (n44) at (-9.84807753012,-1.73648177667){4};
          \node (nb2) at (-9.39692620786,-3.42020143326){B};
          \node (n26) at (-8.66025403784,-5)            {2};
          \node (nc1) at (-7.66044443119,-6.42787609687){C};
          \node (n54) at (-6.42787609687,-7.66044443119){5};
          \node (nc2) at (-5,-8.66025403784)            {C};
          \node (n35) at (-3.42020143326,-9.39692620786){3};
          \node (nd1) at (-1.73648177667,-9.84807753012){D};
          \node (n45) at (0,-10)                        {4};
          \node (nd2) at (1.73648177667,-9.84807753012) {D};
          \node (n36) at (3.42020143326,-9.39692620786) {3};
          \node (ne1) at (5,-8.66025403784)             {E};
          \node (n55) at (6.42787609687,-7.66044443119) {5};
          \node (ne2) at (7.66044443119,-6.42787609687) {E};
          \node (n46) at (8.66025403784,-5)             {4};
          \node (nf1) at (9.39692620786,-3.42020143326) {F};
          \node (n56) at (9.84807753012,-1.73648177667) {5};
          \node (nf2) at (10,0)                         {F};
          
          \foreach \from/\to in {n1/n24,n1/na1,n24/na1,n1/n34,n1/na2,n34/na2,n1/n25,n1/nb1,n25/nb1,n1/n44,n1/nb2,n44/nb2,n1/n26,n1/nc1,n26/nc1,n1/n54,n1/nc2,n54/nc2,n1/n35,n1/nd1,n35/nd1,n1/n45,n1/nd2,n45/nd2,n1/n36,n1/ne1,n36/ne1,n1/n55,n1/ne2,n55/ne2,n1/n46,n1/nf1,n46/nf1,n1/n56,n1/nf2,n56/nf2, n1/n2, n1/n3, n1/n4, n1/n5, n2/n3, n2/n4, n2/n5, n3/n4, n3/n5, n4/n5}
            \draw (\from) -- (\to);
        \end{tikzpicture}
	\caption{A counterexample VI}
	\label{fig:lceVI}
\end{figure}
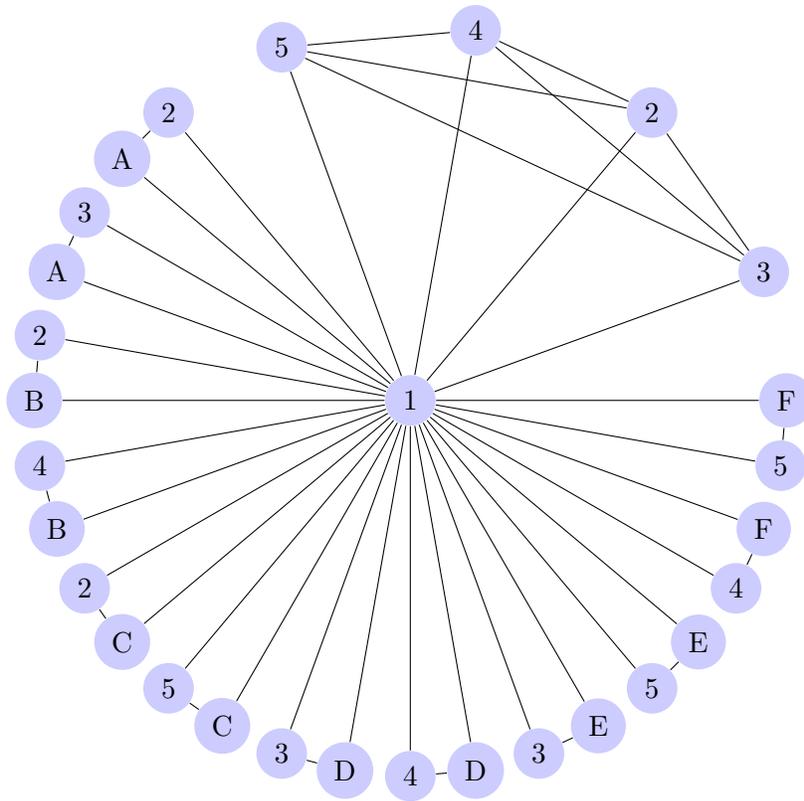

Figure~\ref{fig:lceVI} depicts the correct and desired outcome. However, this is not guaranteed to happen. Alternatively, $\{2, 3\}$ could just as well be merged with $\{2, A\}$ or $\{3, A\}$. In this case, the result of the merge will be the $4$-clique $\{1, 2, 3, A\}$, which is the same clique found in the previous subsection. The algorithm will stop here, since no other $3$-cliques surrounding node 1 could be merged with $\{2, 3, A\}$. Note that the algorithm does not discover the $5$-clique.

\begin{figure}[H]
	\centering
		\begin{tikzpicture}
          [scale=.5,auto=left,every node/.style={circle,fill=blue!20}]
          \node (n1)  at (0,0)                           {1};
          \node (n22) at (8.66025403784,5)              {2};
          \node (n41) at (7.66044443119,6.42787609687)  {4};
          \node (n23) at (6.42787609687,7.66044443119)  {2};
          \node (n51) at (5,8.66025403784)              {5};
          \node (n32) at (3.42020143326,9.39692620786)  {3};
          \node (n42) at (1.73648177667,9.84807753012)  {4};
          \node (n33) at (0,10)                         {3};
          \node (n52) at (-1.73648177667,9.84807753012) {5};
          \node (n43) at (-3.42020143326,9.39692620786) {4};
          \node (n53) at (-5,8.66025403784)             {5};
          \node (n24) at (-6.42787609687,7.66044443119) {2};
          \node (n34) at (-8.19152044289,5.73576436351) {3};
          \node (na2) at (-9.39692620786,3.42020143326) {A};
          \node (n25) at (-9.84807753012,1.73648177667) {2};
          \node (nb1) at (-10,0)                        {B};
          \node (n44) at (-9.84807753012,-1.73648177667){4};
          \node (nb2) at (-9.39692620786,-3.42020143326){B};
          \node (n26) at (-8.66025403784,-5)            {2};
          \node (nc1) at (-7.66044443119,-6.42787609687){C};
          \node (n54) at (-6.42787609687,-7.66044443119){5};
          \node (nc2) at (-5,-8.66025403784)            {C};
          \node (n35) at (-3.42020143326,-9.39692620786){3};
          \node (nd1) at (-1.73648177667,-9.84807753012){D};
          \node (n45) at (0,-10)                        {4};
          \node (nd2) at (1.73648177667,-9.84807753012) {D};
          \node (n36) at (3.42020143326,-9.39692620786) {3};
          \node (ne1) at (5,-8.66025403784)             {E};
          \node (n55) at (6.42787609687,-7.66044443119) {5};
          \node (ne2) at (7.66044443119,-6.42787609687) {E};
          \node (n46) at (8.66025403784,-5)             {4};
          \node (nf1) at (9.39692620786,-3.42020143326) {F};
          \node (n56) at (9.84807753012,-1.73648177667) {5};
          \node (nf2) at (10,0)                         {F};
          
          \foreach \from/\to in {n1/n22,n1/n41,n22/n41,n1/n23,n1/n51,n23/n51,n1/n32,n1/n42,n32/n42,n1/n33,n1/n52,n33/n52,n1/n43,n1/n53,n43/n53,n1/n25,n1/nb1,n25/nb1,n1/n44,n1/nb2,n44/nb2,n1/n26,n1/nc1,n26/nc1,n1/n54,n1/nc2,n54/nc2,n1/n35,n1/nd1,n35/nd1,n1/n45,n1/nd2,n45/nd2,n1/n36,n1/ne1,n36/ne1,n1/n55,n1/ne2,n55/ne2,n1/n46,n1/nf1,n46/nf1,n1/n56,n1/nf2,n56/nf2, n1/n24, n1/n34, n1/na2, n24/n34, n24/na2, na2/n34}
            \draw (\from) -- (\to);
        \end{tikzpicture}
	\caption{A counterexample VII}
	\label{fig:lceVII}
\end{figure}
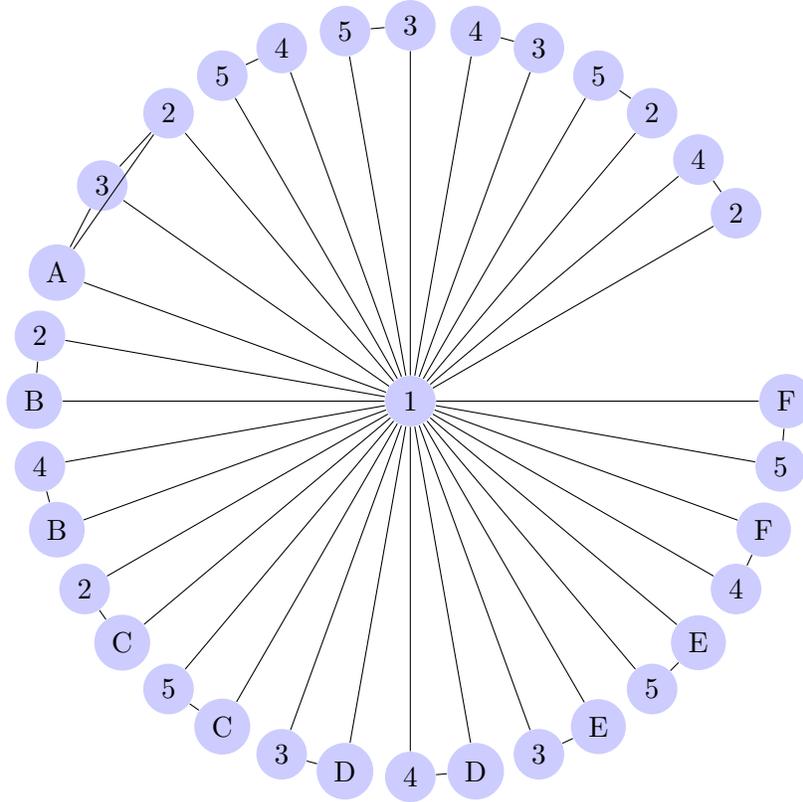

\subsection{Counterexample Conclusion}

Since it is possible that for every pair the algorithm starts a merge with, the $5$-clique is not found, it is also possible that through all iterations LaPlante's algorithm will not find the $5$-clique at all, and thus, will not find the maximum clique in the graph. Therefore, LaPlante's algorithm is incorrect.

This approach to defeating the algorithm works because LaPlante's algorithm begins by finding all $3$-cliques for each vertex. Around each vertex, the algorithm selects each pair first, and tries to merge with that pair. The algorithm can never backtrack from these merges, until it restarts with another vertex-pair as the first merge. Thus if it is possible that each vertex-pair can merge with a vertex-pair not part of the maximum clique, these steps would not be undone and the maximum clique would not be found. The simplest graph of this type that we could find was the situation where the smaller cliques that were ``accidentally'' merged into were of size $4$, and the biggest clique was actually size $5$. From there, all that was necessary was to verify that every combination of $3$ of the $5$ vertices in the max clique was part of a $4$-clique with another extra vertex not part of the $5$-clique. Thus every vertex pair in the $5$-clique would have the possibility of being merged into the $4$-clique instead of the $5$-clique it is in, thus missing the $5$-clique altogether in the search.

\section{Conclusion}

Tamta, Pande, and Dhami and LaPlante both propose polynomial-time algorithms for solving the maximum clique problem, an NP-complete problem. While both algorithms work in many cases, they fail to achieve their claimed behavior. We believe that NP-complete graph algorithms are particularly susceptible to these kinds of proposals, as it's often not obvious how to construct a graph for which a promising heuristic fails. The smallest counterexample we could produce for LaPlante's algorithm has 15 vertices.

After correcting for the trivial errors in Tamta, Pande, and Dhami's paper, both algorithms can be corrected by introducing backtracking. The main counterexamples that we identified all involve forcing the algorithms into positions where they could make unwise choices. Tamta, Pande, and Dhami. state that their algorithm will interdict a vertex with a minimum number of edges connected to it---if there are multiple such vertices, the behavior of the algorithm is unspecified. But if the algorithm is allowed to backtrack through all possible choices, it will have no trouble solving our counterexamples.\footnote{Assuming the trivial errors are dealt with.} LaPlante's algorithm will solve our counterexample under the following condition: in the second part of the algorithm, it may backtrack through all possible choices for a $3$-clique to select during iterations. But in both cases, backtracking voids the inherent polynomial-time efficiency of the algorithms; the backtracking approach may well require exponential time.

\section*{Acknowledgments}
We thank Professor Lane A. Hemaspaandra and teaching assistants Charles Lehner, Adam Scrivener, and Taylan Sen for providing an environment in CSC 200H that encouraged us to collaborate and improve our collective understanding of NP-completeness and clique problems, and for many helpful comments on a draft of this paper. We assume responsibility for any errors that remain.

\nocite{*}
\printbibliography

\end{document}